\documentclass{elsarticle}
\usepackage{graphicx,bm,enumerate}
\usepackage{amssymb,amsmath}
\begin{document}
\title{Analytic form of head-to-head domain walls in thin ferromagnetic cylinders} 
\author{Riccardo Hertel\fnref{cor1}}
\ead{hertel@ipcms.unistra.fr}
\address{Institut de Physique et Chimie des Mat\'eriaux de Strasbourg, Universit\'e de Strasbourg, CNRS UMR 7504, Strasbourg, France} 
\author{Attila K\'akay}  
\address{Peter Gr\"unberg Institut (PGI-6), Forschungszentrum   J\"ulich    GmbH,   D-52428   J\"ulich,   Germany}

\fntext[cor1]{Corresponding author: Riccardo Hertel, Tel.: +33 38810 7263; Fax: +33 38810 7249}
\newcommand{\be}{\begin{equation}}
\newcommand{\ee}{\end{equation}}
\newcommand{\ber}{\begin{eqnarray}}
\newcommand{\eer}{\end{eqnarray}}
\newcommand{\p}{\partial}
\begin{abstract}
The one-dimensional problem of a static head-to-head domain wall structure in a thin soft-magnetic nanowire with circular cross-section is treated within the framework of micromagnetic theory. A radius-dependent analytic form of the domain wall profile is derived by decomposing the magnetostatic energy into a monopolar and a dipolar term. We present a model in which the dipolar term of the magnetostatic energy resulting from the transverse magnetization in the center of the domain wall  is calculated with Osborn's formulas for homogeneously magnetized ellipsoids [Phys.~Rev.~\textbf{67}, 351 (1945)]. The analytic results agree almost perfectly with simulation data as long as the wire diameter is sufficiently small to prevent inhomogeneities of the magnetization along the cross-section. Owing to the recently demonstrated negligible D\"oring mass of these walls, our results should also apply to the dynamic case, where domain walls are driven by spin-transfer toque effects and/or an axial magnetic field. 
\end{abstract}
\maketitle
\section{Introduction}
Head-to-head domain walls in magnetic nanostrips and patterned thin-film elements \cite{mcmichael_head_1997} have received much attention in the past decade because of their potential as units of information in non-volatile memories \cite{parkin_data_2009}, shift registers \cite{hayashi_current-controlled_2008}, and logic devices \cite{allwood_magnetic_2005}. The vast majority of these studies referred to head-to-head domain walls in thin magnetic strips, where depending on the width, the thickness, and on the ferromagnetic material they can occur in two different forms, which have become known as transverse walls and vortex walls \cite{klaui_head--head_2008}. More recently, the attention has shifted from flat strips to magnetic nanocylinders \cite{yan_beating_2010} and nanotubes \cite{yan_chiral_2012,otalora_chirality_2012}. It was predicted \cite{yan_beating_2010} that transverse domain walls in sufficiently thin nanowires with circular cross-section can propagate smoothly, without experiencing the oscillatory behavior that usually occurs above the Walker field \cite{schryer_motion_1974}. 

From a fundamental perspective, it is remarkable that this type of domain wall belongs to a category that is different from the Bloch \cite{bloch_zur_1932} and N\'eel \cite{neel_energie_1955} walls that are usually discussed in the textbooks on magnetism \cite{aharoni_introduction_2000,hubert_magnetic_2012,malozemoff_magnetic_1979}. Head-to-head domain wall structures have been predicted by and intensively studied with micromagnetic simulations, and their properties have been thoroughly investigated in experiments. Nevertheless, they have received less attention concerning their fundamental micromagnetic aspects than the other domain wall types and their mixed forms, in particular concerning analytic theory. In this article we provide an analytic form of the domain wall profile of a transverse head-to-head wall in a thin cylindrical wire. While the overall result is very similar to the usual kink-type transition of Bloch walls or N\'eel walls, the particular distribution of magnetic charges resulting from the geometric confinement requires modifications that are accounted for and discussed in detail by means of a radius-dependent effective demagnetizing factor.

\section{General properties of one-dimensional head-to-head walls}
Because of their occurrence in flat magnetic strips, their in-plane magnetization, and the rotation of the magnetization by 180$^\circ$, transverse head-to-head walls have occasionally been compared or rather misinterpreted in the literature as N\'eel walls or N\'eel type transition regions. But the structure of head-to-head domain walls is very different. The boundary conditions, the orientation of the domain wall with respect to the magnetization in the adjacent domains and, most importantly, the magnetostatic charge distribution is significantly different from that of a N\'eel wall, making such comparisons misleading and inaccurate.
Of all the micromagnetic structures established in the literature, the vertical Bloch line is probably the one that represents closest similarity with a transverse domain wall. The main difference between a vertical Bloch line and a transverse wall in a thin cylindrical wire is the geometric constraint of the latter, an effect that has been predicted to lead to significant deviations in the wall profile \cite{bruno_geometrically_1999}.

The most important static properties of head-to-head walls can be subdivided in three aspects: The orientation of the domain wall, the boundary conditions, and the distribution of magnetic charges.
Contrary to Bloch walls or N\'eel walls, the head-to-head wall is aligned perpendicular to the magnetization in the adjacent domains, which -- as the name suggests -- are oriented in opposite directions. This reflects in the boundary conditions of the magnetization $\lim_{z\to\pm\infty}M_z(z)=\pm M_s$, where $M_z$ is the magnetization component in the $z$-direction, the domain wall is parallel to the $xy$-plane, and $M_s=|\bm{M}|$ is the spontaneous magnetization; a material parameter. In the middle of the domain wall, the magnetization lies in the domain wall plane, in contrast to a N\'eel wall. In spite of these structural differences, it is the charge distribution that represents the most important micromagnetic difference between head-to-head walls and the historically established domain wall types. While the Bloch domain wall is free of stray fields in the idealized case of two infinitely extended half-spaces, the N\'eel wall is characterized by dipolar volume charges, with opposite sign on either side of the domain wall. The head-to-head domain wall, in contrast, has a predominant monopolar magnetostatic field, which is superimposed by a dipolar field with surface charges distributed on opposite sides of the domain wall. This superposition of monopolar and dipolar field contributions is known from vertical Bloch lines \cite{malozemoff_magnetic_1979,hubert_interactions_1974}. In the case of head-to-head walls in thin and flat magnetic strips, this leads to the typical V-shape of the domain wall \cite{thiaville_domain_2002}. Here we investigate a simpler case; a one-dimensional model which does not allow for significant changes of the magnetization perpendicular to the symmetry axis. Such a situation is realized in thin ferromagnetic round wires of negligible anisotropy, which are subdivided into two domains with opposite magnetization, each aligned along the symmetry axis of the wire. Owing to time-inversion symmetry it is not necessary to treat head-to-head domain walls differently from tail-to-tail walls.

\section{Energy functional}

A nanowire can be considered as thin if its diameter is smaller than the width of a domain wall. In these cases, the micromagnetic problem becomes one-dimensional as the magnetization depends -- at least in a good approximation -- only on the position $z$ along the wire, $\bm{M}=\bm{M}(z)$, where the $z$ axis is the symmetry axis of the cylinder. In an infinitely extended wire, the problem of a head-to-head wall is defined with the boundary conditions

\be
\lim_{z\to-\infty}\theta=0, \qquad\lim_{z\to\infty}\theta=\pi,
\ee
where the spherical coordinates $\phi(z)$ and $\theta(z)$ describe the directional cosines of the magnetization, {\em i.e.},
\ber
M_x(z)&=&M_s\sin\phi(z)\cdot\sin\theta(z)\\
M_y(z)&=&M_s\cos\phi(z)\cdot\sin\theta(z)\\
M_z(z)&=&M_s\cos\theta(z)\quad.
\eer
By assuming $d\phi/dz=0$ and using spherical coordinates, the general form of the exchange energy density simplifies significantly:
\be
e_{\rm xc}(z)=A\left(\frac{{\rm d}\theta}{{\rm d}z}\right)^2\quad,
\ee
where $A$ is the exchange constant. If the magnetocrystalline anisotropy is negligible and no external magnetic field is applied, only the energy terms of the magnetostatic energy density $e_{\rm m}$ and the exchange energy density $e_{\rm xc}$ need to be considered. The total energy is
\be
E=r^2\pi\int\limits_{-\infty}^{\infty}e_{\rm xc}(z)+e_{\rm m}(z)\,{\rm d}z
\ee
where $r$ is the radius of the cylindrical wire.
In order to evaluate this integral, an approximate form of $e_{\rm m}(z)$ is necessary.

\section{Domain wall profile and width}
The monopolar contribution of the magnetostatic field is largely determined by the boundary conditions. By means of the boundary conditions, the value of the magnetostatic volume charges $\rho=-\bm{\nabla M}$ can immediately be determined to $\rho=2\pi r^2M_{\rm s}$, which neutralizes the magnetostatic surface charges $\sigma_e=-\pi r^2M_{\rm s}$ at each end of the wire. Even though in the domain wall region the monopolar field is the dominant magnetostatic term, it is discarded from the rest of the analysis. The reasoning behind this is that the monopolar term provides an energetic offset that does not change significantly with the domain wall width. This is in contrast to the dipolar magnetostatic term associated with the transverse component of the magnetization. Unlike the constant value of the volume charges $\rho$, the amount of surface charges $\sigma_{\rm w}$ and hence the demagnetizing field generated by those charges depends strongly on the width of the domain wall.

We assume that the dipolar magnetic field can be described by an effective demagnetizing factor $N_\phi$
\be
E=r^2\pi\int\limits_{-\infty}^{\infty}\left[A\left(\frac{{\rm d}\theta}{{\rm d}z}\right)^2+\frac{\mu_0M_s^2}{2}N_\phi\sin^2\theta\right]\,{\rm d}z\label{nrg}
\ee
where $\mu_0=4\pi\cdot10^{-7}$Vs/Am is the vacuum permeability.

The standard formalism known for one-dimensional domain walls can be employed, {\em i.e.}, minimizing the energy functional (\ref{nrg}) with respect to $\theta$ \cite{bloch_zur_1932,aharoni_introduction_2000,hubert_magnetic_2012,landau_theory_1935}. The solution of the variational problem $\delta E=0$ yields
\be\label{profile}
\cos\theta=\tanh (z/\xi);\qquad\xi=\sqrt{\frac{2A}{\mu_0N_\phi M_s^2}}\qquad
\ee
According to Lilley's definition \cite{hubert_magnetic_2012,lilley_energies_1950}, the domain wall width $\Delta$ is 
\be
\Delta=\pi\sqrt{\frac{2A}{\mu_0N_\phi M_s^2}}\qquad.
\ee

The center of the wall is characterized by $\theta=\pi/2$ and is located at $z=0$. In the vicinity of the center of the domain wall one can therefore use the approximation $\cos\theta=-\sin(\theta-\pi/2)\simeq-\theta$.
The derivative $({\rm d}\theta/{\rm d}z)$ at the center $c$ of the domain wall is therefore
\ber
\left.\frac{{\rm d}\theta}{{\rm d}z}\right|_{c}&=&-\left.\frac{{\rm d}}{{\rm d}z}\tanh (z/\xi)\right|_c\\
&=&-\left.\frac{\xi}{\xi^2-z^2}\right|_c\\
&=&-\sqrt{\frac{\mu_0N_\phi M^2_s}{2A}} \label{final}
\eer
The value of the first derivative at the center is important for the mobility of the domain wall \cite{yan_beating_2010}.
It was recently demonstrated \cite{yan_beating_2010} that this type of domain wall has an almost vanishing D\"oring mass \cite{doring_uber_1948}, meaning that the profile of the domain wall hardly changes when it is in motion. Hence, the zero-field solution for the derivative $({\rm d}\theta/{\rm d}z)|_c$ according to eq.~(\ref{final}) should hold also in the dynamic case.

\section{Effective demagnetizing factors\label{main}}
The demagnetizing factor $N_\phi$ of an infinitely extended round cylinder \cite{aharoni_introduction_2000,hubert_magnetic_2012} is equal to $1/2$ . While our geometry of a thin and long wire with round cross-section corresponds very well to this limiting case, this value is an inappropriate estimate in the case of a head-to-head wall. The demagnetizing factor $N_\phi=1/2$ applies to a cylindrical wire with homogeneous magnetization perpendicular to the symmetry axis. In contrast to this, the magnetization in a wire with a head-to-head domain wall structure is almost everywhere aligned with the axis. Only the domain wall region provides magnetostatic surface charges $\sigma_w$ on the barrel of the cylinder. Hence, the value of the effective demagnetizing factor is considerably smaller than $1/2$. 

\begin{figure}
\centerline{\includegraphics[width=7cm]{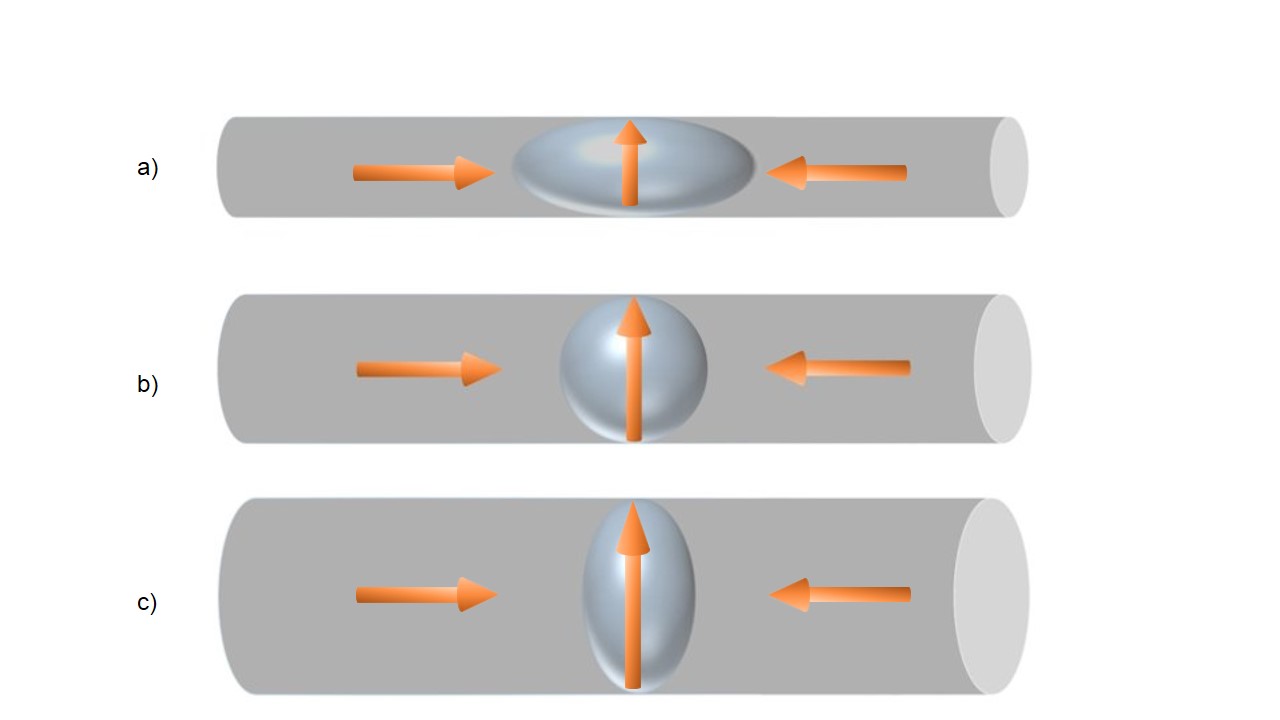}}
\caption{Only the region magnetized perpendicular to the wire axis provides an adjustable contribution to the magnetostatic energy. In a thin soft-magnetic wire of diameter $d=\pi l_s$ this region can be approximated accurately by assuming a sphere (b).\label{h2h-sketch} If the wire is thinner, the sphere becomes a prolate spheroid (a), and an oblate spheroid in thicker wires (c).}
\end{figure}

To obtain a useful estimate for the effective demagnetizing factor we use a model as shown in Fig.~\ref{h2h-sketch}: The dipolar magnetostatic contribution in the head-to-head wall is assumed to originate from a region that is approximated with a homogeneously magnetized spheroid. The symmetry axis of the spheroid coincides with the wire axis, and the magnetization in the spheroid is perpendicular to the wire. The advantage of using this model consists in the analytic forms that are available for demagnetizing factors of homogeneously magnetized ellipsoids.
According to Osborn \cite{osborn_demagnetizing_1945} the demagnetizing factor of a spheroid magnetized homogeneously perpendicular to its symmetry axis is given by the following two equations. 
A prolate spheroid with semi-axes $a, b, c$ ($b=c$, $a>b$), $m=a/b$ yields a demagnetizing factor 
\be\label{N-prol}
N^{(p)}_\phi=\frac{m}{2(m^2-1)}\left[m-\frac{1}{2\sqrt{m^2-1}}\ln\left(\frac{m+\sqrt{m^2-1}}{m-\sqrt{m^2-1}}\right)\right]
\ee
In the case of an oblate spheroid ($b=c$, $a<b$, $m=b/a$), one has
\be\label{N-obl}
N^{(o)}_\phi=\frac{1}{2(m^2-1)}\left[\frac{m^2}{\sqrt{m^2-1}}\cdot\arcsin\left(\frac{\sqrt{m^2-1}}{m}\right)-1\right]
\ee
Note the different definition of $m$ in equations (\ref{N-prol}) and (\ref{N-obl}) which ensures $m>1$ in both cases.

We assume that the shape of the spheroid is a function of the radius of the wire, expressed in units of the magnetostatic exchange length $l_s=\sqrt{2A/(\mu_0M_s^2)}$. At a specific wire radius $r_0$ the spheroid has the shape of sphere, in thinner wires the domain wall region is approximated as a prolate spheroid, and in thicker wires as an oblate spheroid.

The model sketched in Fig.~\ref{h2h-sketch} contains two aspects that are {\em a priori} unknown, and which can be calibrated, {\em e.g.}, by comparison with numerical data: The value $r_0$ and the functional dependence $m(r)$ must be defined, where $m$ is the largest ratio of the pairs of semi-axes and $r_0$ is the wire radius at which the ellipsoid becomes spherical. Over a rather broad range of diameters, we obtain good agreement between analytic results and numerical simulations by setting $r_0=l_s\cdot\pi/2$ and assuming that the largest ratio $m$ of the half-axes of the ellipsoid is equal to $\max\{r_0/r;r/r_0\}$. Since the two half-axes perpendicular to the wire $b$ and $c$ are identical to the wire radius $r$, the missing half-axis $a$ along the wire is determined according to $a=r_0$ if $r<r_0$ and $a=r^2/r_0$ if $r>r_0$, where $r_0=l_s\pi/2$. Hence, for thin wires ($r<r_0$)  one obtains 
\be\label{mr0}
m(r)=\frac{r_0}{r}
=\frac{\pi}{2}\sqrt{\frac{2A}{\mu_0M_s^2}}\cdot\frac{1}{r}
\ee
and in the case $r>r_0$, correspondingly,
\be\label{mr}
m(r)=\frac{r}{r_0}
=\frac{2}{\pi}\sqrt{\frac{\mu_0M_s^2}{2A}}\cdot r\quad.
\ee
The values of $m$ can be inserted in Eqs.~(\ref{N-prol}), (\ref{N-obl}) to calculate $N_\phi(r)$. Once the value of $N_\phi(r)$ is determined, Eq.~(\ref{profile}) yields the radius-dependent profile of the head-to-head domain.

The numerical simulations used to calibrate the analytic model and to calculate the domain wall profile were performed with our general-purpose finite-element micromagnetic code {\tt TetraMag} \cite{kakay_speedup_2010,hertel_guided_2007}, the same code which was also used to calculate the field- and current-driven magnetization dynamics of these domain walls in nanowires \cite{yan_beating_2010}.

\section{Comparison with simulation Results}
\begin{figure}
\centerline{\includegraphics[width=7cm]{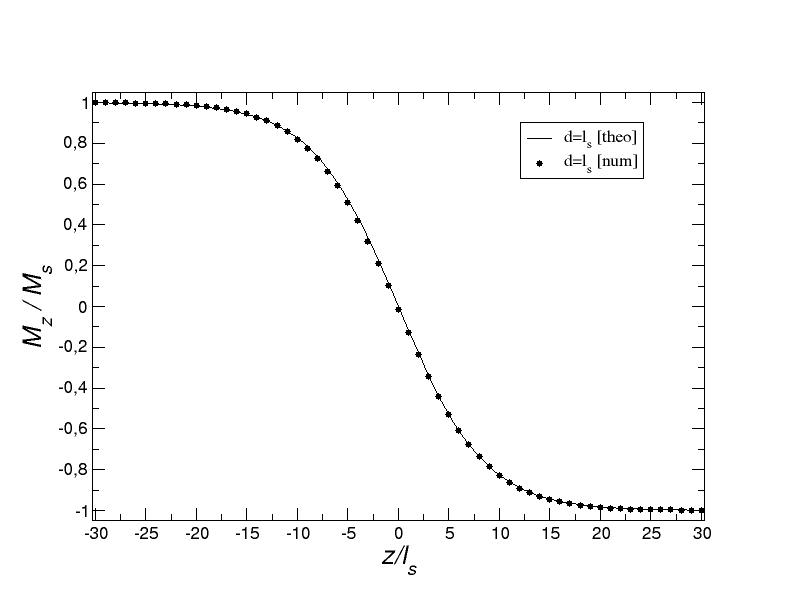}}
\caption{Comparison between analytic data and simulation results. The continuous line displays the magnetization profile $M_z(z)/M_s$ calculated analytically with the model described in the text. The open dots represent results from a finite-element simulation. The data refers to a head-to-head wall in a very thin cylindrical wire with diameter $d=2r=l_s$.  \label{profile-fig}}
\end{figure}
As shown in Fig.~(\ref{profile-fig}), the model described in the previous section yields an almost perfect agreement with simulation data in the case of very thin wires. Deviations become more pronounced as the diameter increases. 

\begin{figure}
\centerline{\includegraphics[width=7cm]{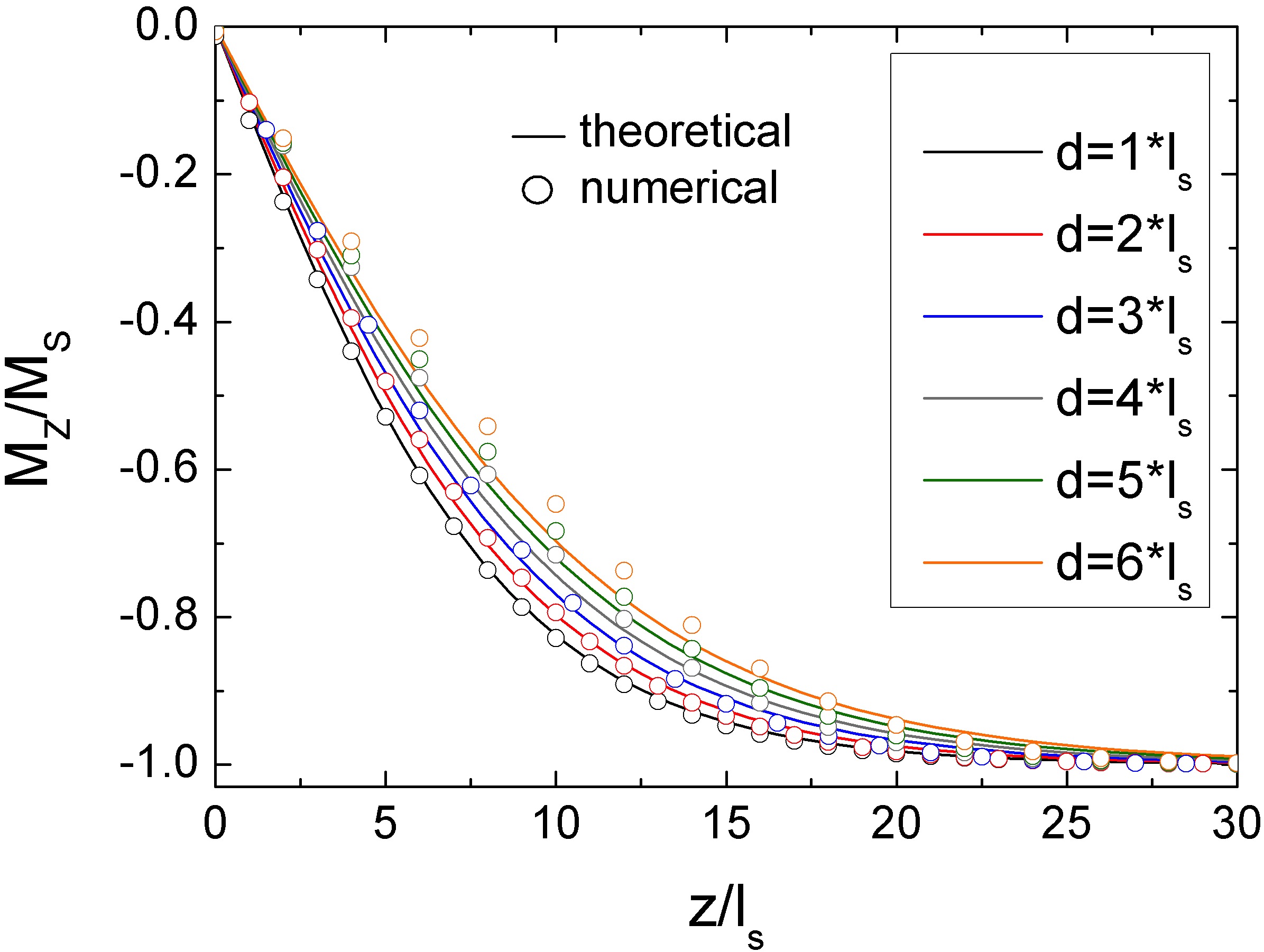}}
\caption{Computed data and analytic results for different wire thicknesses. For symmetry reasons it is sufficient to display only the region $z\ge0$. The scattered open dots represent simulation results and the lines are analytic profiles, as noted in the figure. The comparison shows data for six wire thicknesses, at equidistant levels ranging from $d=l_s$ to $d=6\cdot l_s$. The numerical data is very well reproduced for wire diameters smaller than $4l_s$. For larger wires deviations occur which are attributed to the three-dimensional structure developing in the domain wall region.\label{comparison2}}
\end{figure}

\begin{figure}
\centerline{\includegraphics[width=7cm]{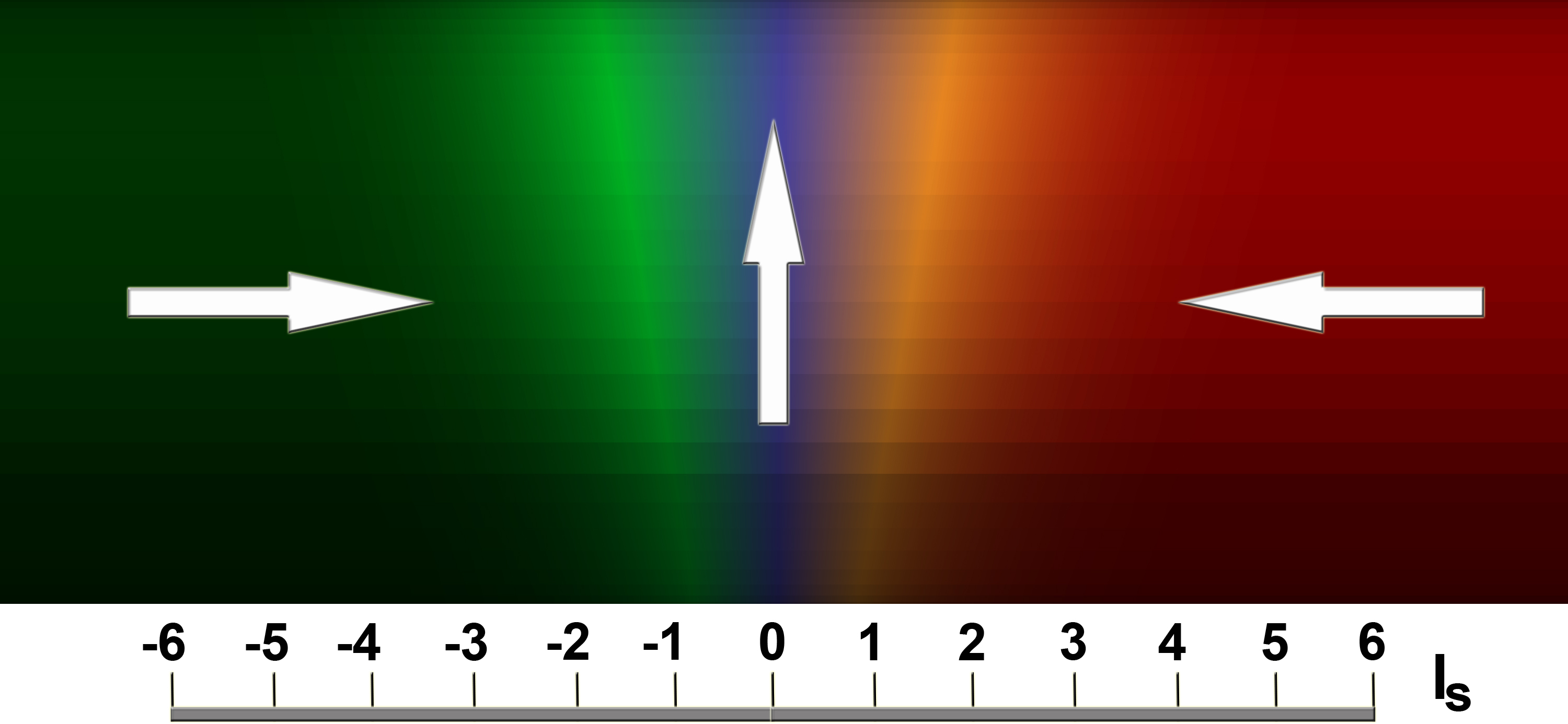}}
\caption{Simulated head-to-head-wall transition in a thick soft-magnetic wire of diameter $d=6\cdot l_s$ The color coding, ranging from green to red, displays the $z$-component of the magnetization, the scale on the bottom shows the length in units of the exchange length $l_s$. The combined effect of the monopolar and dipolar magnetostatic charges in the transition region leads to the typical ``V-shape" known from thin-film elements: The domain wall is broader on one side of the wire than the other. This illustrates the breakdown of a fundamental assumption of our model, {\em i.e.}, that the wire is thin enough so that the magnetization only depends on $z$. In view of this qualitative change in the domain wall structure, the deviations displayed in Fig.~\ref{comparison2} for thick wires are not surprising.\label{thick}}
\end{figure}

In the case of thicker wires, with a diameter of more than three times the exchange length, the domain wall width and the slope at center of the domain wall is still well reproduced by the model, but the profile of the magnetization no longer follows precisely the kink described by the $\tanh$ curve. This indicates that the thin-wire approximation, according to which the magnetization depends only on the position $z$, reaches its limits of validity. The computed magnetic structure displayed in Fig.~(\ref{thick}) shows how the aforementioned V-shaped transition unfolds in a wire of a diameter $d=6\cdot l_s$.

\section{Linearization}
Owing to the calibration with numerical results and the assumptions described in section \ref{main} we have obtained a purely analytic model that holds for any soft-magnetic thin nanowire with circular cross-section. Evaluating Osborn's equations (\ref{N-prol}), (\ref{N-obl}) can however be quite tedious if one is only interested in a quick estimate, {\em e.g.}, of the domain wall width. 
To simplify the numerical evaluation we provide here a Taylor expansion of the terms $(N_\phi^{(o)})^{1/2}$  and $(N_\phi^{(p)})^{1/2}$. The expansion is performed around $m=1$, so that $\chi=m-1$ is the small parameter describing the deviation of the demagnetizing ellipsoid from a spherical shape. 

In the case $r>r_0$ the spheroid is oblate and $m=r/a$, $r>a$. If $r\simeq r_0$ the square root of the demagnetizing factor can be approximated as
\be\label{tay1}
\sqrt{N_\phi^{(o)}}(\chi)=\frac{1}{\sqrt{3}}-\frac{1}{5\sqrt{3}}\chi+\frac{3\sqrt{3}}{175}\chi^2+\mathcal{O}(\chi^3)\quad,
\ee
while for prolate spheroids $m=a/r$, $a>r$ the series 
\be\label{tay2}
\sqrt{N_\phi^{(p)}}(\chi)=\frac{1}{\sqrt{3}}+\frac{1}{5\sqrt{3}}\chi-\frac{26}{175\sqrt{3}}\chi^2+\mathcal{O}(\chi^3)\quad
\ee
can be used. To apply these approximations, $m(r)$ can be calculated according to eq.~(\ref{mr0}) or eq.~(\ref{mr}) in order to determine $\chi=m-1$. If $\chi\ll1$ the above Taylor series should represent good estimates.

\section{Conclusion}
We have provided a radius-dependent analytic form of the head-to-head domain wall in thin cylindrical soft-magnetic wires. Osborn's formulas \cite{osborn_demagnetizing_1945} for homogeneously magnetized ellipsoids have been used to obtain estimates for the effective demagnetizing factor in the domain wall transition region. Discarding the predominant monopolar magnetostatic contribution of the head-to-head domain wall as an energetic offset has proven to be a suitable assumption. The analytic values of the domain wall profile compare very well with the numerical ones, as long as the assumption of a one-dimensional transition is valid, {\em i.e.}, if the magnetization does not show any significant radial dependence. 
The model depends on various assumptions which are incorporated in Eqs.~(\ref{mr0}) and (\ref{mr}). Other forms of $m(r)$ are possible, which may lead to even better results.
The analytic form of the head-to-head domain wall profile described in this article allows to calculate the derivative of the magnetization at the center of the domain wall according to eq.~(\ref{final}), which is an essential parameter for the mobility of these domain walls \cite{yan_beating_2010}. In many cases a Taylor series [Eqs.~(\ref{tay1}), (\ref{tay2})] can be used to approximate the complicated equations for the square root of the demagnetizing factor.

\bibliographystyle{elsarticle-num}

\end{document}